\documentstyle[12pt,epsf,axodraw,rotate]{article}

\begin{document}

\parskip=0.3cm
\begin{titlepage}

\hfill \vbox{\hbox{DFPD 00/TH/08}\hbox{UNICAL-TH 00/1}\hbox{BITP-00-05E}
\hbox{February 2000}}

\vskip 0.2cm

\centerline{\bf THE POMERON AS A FINITE SUM OF GLUON LADDERS $~^\diamond$}

\vskip 0.5cm

\centerline{R.~Fiore$^{a\dagger}$, L.L.~Jenkovszky$^{b\S}$,
A.~Lengyel$^{c\S}$, F.~Paccanoni$^{d\ast}$, A.~Papa$^{a\dagger}$}

\vskip 0.1cm

\centerline{$^{a}$ \sl  Dipartimento di Fisica, Universit\`a della Calabria,}
\centerline{\sl Istituto Nazionale di Fisica Nucleare, Gruppo collegato di Cosenza}
\centerline{\sl I-87036 Arcavacata di Rende, Cosenza, Italy}
\vskip .15cm
\centerline{$^{b}$ \sl  Bogoliubov Institute for Theoretical Physics,}
\centerline{\sl Academy of Sciences of the Ukraine}
\centerline{\sl 252143 Kiev, Ukraine}
\vskip .15cm
\centerline{$^{c}$ \sl Institute of Electron Physics, }
\centerline{\sl v. Universitetska 21, 88000 Uzhgorod, Ukraine}
\vskip .15cm
\centerline{$^{d}$ \sl Dipartimento di Fisica, Universit\`a di Padova,}
\centerline{\sl Istituto Nazionale di Fisica Nucleare, Sezione di Padova}
\centerline{\sl via F. Marzolo 8, I-35131 Padova, Italy}

\vskip 0.1cm

\begin{abstract}
A model for the Pomeron at $t=0$ is suggested. It is based on the idea of a
finite sum of ladder diagrams in QCD. Accordingly, the number of $s$-channel
gluon rungs and correspondingly the powers of logarithms in the forward
scattering amplitude depends on the phase space (energy) available, i.e. as
energy increases, progressively new prongs with additional gluon
rungs in the $s$-channel open. Explicit expressions for the total
cross section involving two and three rungs or, alternatively, three and
four prongs (with $\ln^2(s)$ and $\ln^3(s)$ as highest terms) is fitted to the
proton-proton and proton-antiproton total cross section data in the
accelerator region. 

\end{abstract}

\vskip 0.2cm

\hrule

\vskip 0.1cm

\noindent
$^{\ast}${\it Work supported by the Ministero italiano dell'Universit\`a e 
della Ricerca Scientifica e Tecnologica and by the INTAS.}
\vfill
$
\begin{array}{ll}
^{\dagger}\mbox{{\it e-mail address:}} &
   \mbox{FIORE,~PAPA@CS.INFN.IT} \\
^{\ddagger}\mbox{{\it e-mail address:}} &
\mbox{JENK@BITP.KIEV.UA} \\
  ^{\S}\mbox{{\it e-mail address:}} &
   \mbox{SASHA@LEN.UZHGOROD.UA} \\
 ^{\ast}\mbox{{\it e-mail address:}} &
   \mbox{PACCANONI@PD.INFN.IT}
\end{array}
$

\vfill
\end{titlepage}
\eject
\textheight 210mm \topmargin 2mm \baselineskip=24pt

\section{Introduction}

It is widely accepted that the Pomeron in QCD corresponds to an infinite
gluon ladder with Reggeized gluons on the vertical lines (see Fig.~1), 
resulting~\cite{FKL,BL,L} in the so-called supercritical behavior 
$\sigma _t\sim s^{\alpha (0)}$, where $\alpha (0)$ is the intercept of the 
Pomeron trajectory. However, at finite energies only a finite number of 
diagrams contributes. The lowest order diagram is that of two-gluon exchange, 
first considered by Low and
Nussinov~\cite{LN}. The next order, involving an $s$-channel gluon rung,
was studied e.g. in the papers~\cite{BL,McCoy}. The problem
of calculating these diagrams is twofold. One problem is connected with the
non-perturbative contributions to the scattering amplitude in the
``soft'' region. It may be ignored by ``freezing'' the running coupling
constant at some fixed value of the momentum transfer and assuming that the
forward amplitude can be cast by a smooth interpolation to $t=0$. More
consistently, one introduces a non-perturbative model~\cite{Francesco} of the
gluon propagator valid also in the forward direction. The second problem is
more technical: at any given perturbative order $\alpha_s^n$, the leading 
contribution in the $s\to\infty$ limit, proportional to $(\alpha_s \ln(s))^n$, 
is given by 
a subset of all the Feynman diagrams contributing at that perturbative order; 
each of these diagrams consists of a leading term in the 
$s\to\infty$ limit and of a non-leading, negligible part. The leading contributions
from all orders in perturbation theory can be resummed~\cite{FKL,BL,L}.
For non-asymptotic energies, however, at any order in the coupling constant
subleading terms are present coming both from the neglected
diagrams and from the neglected part of the leading diagrams. 
Although functionally the result is
always the sum of increasing powers of logarithms, the numerical values of
the coefficients entering the sum is lost unless all diagrams are calculated.

The summation and convergence of an infinite series is a known problem in
physics. As discussed in a recent paper~\cite{West}, various situations may
occur, where a finite series approximates the exact result better than the
infinite sum does. Since, as stressed above, the coefficients of the
perturbative series are not known from QCD even for $t=0$ (their calculation
in the non-forward direction is much more tricky), the convergence of the series
is also unknown.

Conversely, one can expand the "supercritical" Pomeron $\sim s^{\alpha(0)}$
in powers of $\ln(s)$. Such an expansion is legitimate within the range of
active accelerators, i.e. near and below the TeV energy region, where fits
to total cross sections by a power or logarithms are known~\cite{Ezhela} to
be equivalent numerically. Moreover, forward scattering data (total cross
sections and the ratio of the real to the imaginary part of the forward
scattering amplitude) do not discriminate even between a single and
quadratic fit in $\ln(s)$ to the data.

Phenomenologically, more information on the nature of the series can be
gained if the $t$ dependence is also involved. The well-known (diffractive)
dip-bump structure of the differential cross section can be roughly imitated
by the Glauber (or eikonal) series, although more refined studies within the dipole
Pomeron model (DP) (linear behavior in $\ln(s)$)~\cite{JShS} show that the
relevant series is not just the Glauber (eikonal) one. A generalization of the DP
model including higher terms in $\ln(s)$ was considered in~\cite{Pierre}. We
mention these attempts only for the sake of completeness, although we stick
to the simplest case of $t=0$, where there are hopes to have some connection
with the QCD calculations.

In the present paper we consider a new parametrization for total cross
sections based on the contribution of a finite series of QCD diagrams with
relative weights (coefficients) and rapidity gaps to be determined from the
data. Each set of the diagrams is "active" in "its zone", i.e. the
parameters should be fitted in each energy interval separately and the
relevant solutions should match. The matching procedure will be similar to
that known for the wave functions in quantum mechanics, i.e. we require
continuity of the total cross section and of its first derivative.

\section{Description of the model and an example with two gluon rungs}

The Pomeron contribution to the total cross section is represented in the form
\begin{equation}
\sigma^P(s)=\sum_{i=0}^N f_i(s) \:\theta(s-s_0^i)\:\theta(s_0^{i+1}-s)\;,
\label{sigma}
\end{equation}
where
\begin{equation}
f_i(s)=\sum^i_{j=0}a_{ij}L^j\;,
\end{equation}
$s_0$ is the prong threshold, $\theta(x)$ is the step function and 
$L\equiv \ln(s)$. Here and in the following, 
for $s$ and $s_0$ it is understood $s/(1\:\mathrm{GeV}^2)$ and 
$s_0/(1\:\mathrm{GeV}^2)$, respectively. 
The main assumption in Eq.~(1) is that the widths of the rapidity gaps 
$\ln(s_0)$ are the same along the ladder and are energy independent. Their
magnitude is not known {\it a priori}, but can be related to correlations between
jets or multiclusters~\cite{BP} formed by single gluons or determined
empirically. The functions $f_i(s)$ are finite polynomials in $L$, 
corresponding to finite gluon ladder
diagrams in QCD, where each power of the logarithm collects all the relevant
diagrams. Each time the rapidity gap $\sim \ln(s)$ exceeds the threshold
value $\ln (s_0)$, a new prong opens adding a new power in $L$.

In Eq.~(1) the sum over $N$ is a finite one, since $N$ is proportional to 
$\ln(s)$, where $s$ is the present squared c.m. energy. Hence this model 
is quite different from the usual approach where, in the limit $s \rightarrow
\infty$, the infinite sum of the leading logarithmic contributions gives 
rise to an integral equation for the amplitude.

To make the idea more clear, we first describe the mechanism in the case of
three gaps (two rungs) with
\begin{equation}
\label{f3}f_0(s)=a_{00}\;,
\end{equation}
\begin{equation}
f_1(s)=a_{10}+a_{11}L\;,
\end{equation}
\begin{equation}
f_2(s)=a_{20}+a_{21}L+a_{22}L^2\;.
\end{equation}
By imposing the requirement of continuity (of the cross section and of its
first derivative) one constrains the parameters. E.g., from the equality $
f_1(s_0)=f_0(s_0)$ the relation
\begin{equation}
a_{10}=a_{00}-a_{11}\ln (s_0)
\end{equation}
follows. Furthermore, from $f_2(s_0^2)=f_1(s_0^2)$ one gets
\begin{equation}
a_{20}=-2a_{21}\ln (s_0)-4a_{22}\ln^2(s_0)+a_{10}+2a_{11}\ln(s_0)
\end{equation}
and from the continuity of the relevant derivatives
\begin{equation}
a_{21}=a_{11}-4a_{22}\ln(s_0)
\end{equation}
follows.

To remedy the effect of the opening prongs and get a smooth behavior at low
energies, we have included also a Pomeron daughter, behaving like $\sim 1/s$, to
the Eqs.~(3) and (4) with parameters $b_0$ and $b_1$
respectively (otherwise the continuity condition could not be applied to
the first, constant term, whose derivative vanishes).

In fitting the model to the data, we rely mainly on $\bar pp$ data that
extend to the highest (accelerator) energies, to which the Pomeron is
particularly sensitive. To increase the confidence level, $pp$ data were
included in the fit as well. To keep the number of free parameters as
small as possible and following the successful phenomenological approach of
Donnachie and Landshoff~\cite{DL}, a single ``effective'' Reggeon trajectory 
with intercept $\alpha \left( 0\right) $ will account for non-leading 
contributions, thus leading to the following form for the total
cross section:
\begin{equation}
\sigma_t(s)=\sigma^P(s)+R_{}^{}(s)\;,
\end{equation}
where $\sigma^P(s)$ is given by Eq.~(\ref{sigma}) and 
$R_{}^{}(s)=a_{}^{}s^{\alpha (0)-1}$ (the parameter $a$ is different 
for $\overline{p}p$ and $pp$ and is considered as an additional free parameter).

Ideally, one would let free the width of the gap $s_0$ and consequently the
number of gluon rungs (highest power of $L$). Although possible, technically
this is very difficult. Therefore we proceed by trial and error, i.e. make
fits for fixed (two and three) number of rungs (power of the logarithms). Even
within this approximation there is some room, as we shall see, to study the $
s_0$ dependence of the results.

Notice that the values of the parameters depend on the energy range of the
fitting procedure. For example, the values of the parameters in $f_0$ if
fitted in "its" range, i.e. for $s\leq s_0$, will get modified with the
higher energy data and correspondingly higher order diagrams included.

Fits to the $\bar pp$ and $pp$ data were performed from $\sqrt{s}=4$ up
to the highest energy Tevatron data (for $\bar pp$), including all the results
from there~\cite{data}. To cover the energy range with equal rapidity gaps 
uniformly, $s_0$ was chosen to be equal to 144.

The resulting fit is shown in Fig.~2. The values of the fitted parameters
are quoted in Table 1.

\section{Three gluon rungs and fits to the $p\bar p$ and $pp$ data}

We cover the energy span available in the accelerator
region by four gaps resulting in three gluon rungs and consequently $L^3$ as the
maximal power. The individual Pomeron terms and relevant gaps now are
\begin{eqnarray}
4\geq \sqrt{s} \leq \sqrt s_0 \;\;\; &:&  f_0  = a_{00}+b_0/s \;, \\
\sqrt s_0  \geq \sqrt{s} \leq (\sqrt s_0)^2 \;\;\; &:& f_1  
= a_{10}+b_1/s+a_{11}L \;,  \\
(\sqrt s_0)^2\geq \sqrt{s} \leq (\sqrt s_0)^3 \;\;\; &:& f_2  
= a_{20}+a_{21}L+a_{22}L^2 \;, \\
(\sqrt s_0)^3\geq \sqrt{s} \leq (\sqrt s_0)^4 \;\;\; &:& f_3  
= a_{30}+a_{31}L+a_{32}L^2+a_{33}L^3. 
\end{eqnarray}

As in the case of three gaps, the individual Pomeron terms and their derivatives
were matched at the prong values. E.g. $a_{10}$ was determined from the
condition $f_0(s_0)=f_1(s_0)$, while from the equality of the relevant
derivatives, $b_1$ can be expressed in terms of the other parameters.
Ultimately, we are left with nine free parameters: $a_{00}$, $b_0$, $a_{11}$,
$a_{22}$, $a_{32}$ and $a_{33}$, each determined in its range,
while $a$ and $\alpha \left( 0\right) $ are fitted in the whole range of the
data. The parameter $s_0$ in principle is also free, but as discussed
above, we determine it by trial and error, starting with the value $s_0=64$.
The gap width was chosen such as to cover the whole rapidity span by at most
four gaps (to have $L^3$). The final value for $s_0$ turned out to be 42.5, 
resulting in a sequence of energy intervals ending at $\sqrt s$ = 1800. 

Fig.~3 shows our fit to the $p\bar p$ and $pp$ total cross section data. The
values of the fitted parameters are quoted in Table 2. The value of the $\chi
^2/$d.o.f. is $\sim 1.3$, much better than in the case of two gluon rungs.

The value of the effective Reggeon intercept remains rather low, close to
0.5 (compare for example with Ref.~\cite{Kaidalov}), however interestingly enough 
its value is correlated with the gap width: the smaller the gap width, 
the higher the Reggeon intercept.

\section{Conclusions}

Although high quality fits were not the primary goal of the present study,
we may conclude that our results are comparable with those of 
similar analyses~\cite{DL}.
There is still room for some technical improvements in this
direction. Our main goal instead was to seek for a correct form of the
``perturbative'' series of total cross sections and for regularities in the
behavior of the parameters. In fact we find that the coefficients in front
of leading logarithms in the Pomeron contribution are related roughly by a
factor 1/10. Notice the alternating signs in front of the logarithms. 
They may reflect the fact discussed in the introduction,
namely that each power of the logarithms collects various contributions of
the same order but from different diagrams (see Fig.~1).

``Footprints'' of the prongs at low energies are slightly visible in Fig.~2
(especially in the case of $pp$ scattering where the contribution 
from secondary Reggeons is smaller than in $\bar pp$). A more detailed study of this
phenomenon could answer the question whether this is an artifact or a
manifestation of the Pomeron's basic properties.

The present model and its experimental verification may shed light also on
the energy range of the applicability of various approximations to the
Pomeron. The simplest, Low-Nussinov model with constant cross sections is a
crude approximation to reality. The inclusion of one gluon rung may be
associated with the dipole Pomeron. This model has many attractive features,
such as selfconsistency with respect to $s$-channel unitarity. Note that 
$\ln(s)$ is the strongest rise within the Regge pole model. The next order,
$\ln^2(s)$, conflicts with the unitarity bound, requiring that the rise
of the cross section does not exceed that of the slope parameter (shrinkage
of the cone), that in the Regge approach is at most $\ln(s)$ (unless special
assumptions are involved) . This regime seems typical of the Tevatron energy
region. The role and weight of higher order terms is interesting, but needs
more care for two reasons: first, too many free parameters make their
determination difficult and secondly, they violate the Froissart bound,
therefore eikonal corrections - otherwise present everywhere - here become
crucial. A generalization of the above procedure within the eikonal model is
possible, although the calculations (matching, fitting) become more
complicated.

Extrapolations to still higher energies are of great interest. By fitting
the model to the cosmic ray data and/or future (RHIC, LHC) accelerator data,
one could explore the role of the new thresholds. On the other hand, from the
present fits and, hopefully, from the QCD calculations one may try to deduce
recursion relations and try to extrapolate the value of the coefficients in
front of the logarithms. In any case, the higher the energy, the more
important the unitarity corrections will become.
Future fits of the model to various data may settle some details left open
by this paper.

\section{Acknowledgment}

We thank V.S.~Fadin, E.A.~Kuraev and L.N.~Lipatov for numerous discussions on the
BFKL Pomeron.

\newpage

\centerline{\bf TABLES}

\vspace{3cm}

\begin{table}[h]
\centering
\begin{tabular}{||c|c|c|c|c|c||}
\hline
$s_0$ & $b_0$ & $b_1$ & $\alpha \left( 0\right)$ & $a_{p\bar p}^{}$ & $a_{pp}^{}$ \\
\hline
144 & -74.4 & 421 & 0.421 & 127 & 45.9 \\
\hline
\hline
$a_{00}$ & $a_{10}$ & $a_{11}$ & $a_{20}$ & $a_{21}$ & $a_{22}$ \\
\hline
35.7 & 15.2 & 3.44 & 57.3 & -5.03 & 0.427 \\
\hline
\end{tabular}
\caption{Fitted parameters in the case of two rungs. The parameters $b_i$ and 
$a_{\ldots}$ are given in units of 1 mb.}
\end{table}

\vspace{2cm}

\begin{table}[h]
\centering
\begin{tabular}{||c|c|c|c|c|c||}
\hline
$s_0$ & $b_0$ & $b_1$ & $\alpha(0)$ & $a_{\bar pp}^{}$ & $a_{pp}^{}$ \\
\hline
42.5 & -89.3 & 43.0 & 0.550 & 105 & 56.3 \\
\hline
\hline
$a_{00}$ & $a_{10}$ & $a_{11}$ & $a_{20}$ & $a_{21}$ & $a_{22}$ \\
\hline
30.6 & 15.5 & 3.22 & 27.3 & 0.0629 & 0.211 \\
\hline
\hline
$a_{30}$ & $a_{31}$ & $a_{32}$ & $a_{33}$ & & \\
\hline
106. & -18.7 & 1.68 & -0.0376 & & \\
\hline
\end{tabular}
\caption{Fitted parameters in the case of three rungs. The parameters $b_i$ and 
$a_{\ldots}$ are given in units of 1 mb. The value of the $\chi^2/$d.o.f.
is $\sim 1.3$.}
\end{table}

\newpage

\phantom{.}

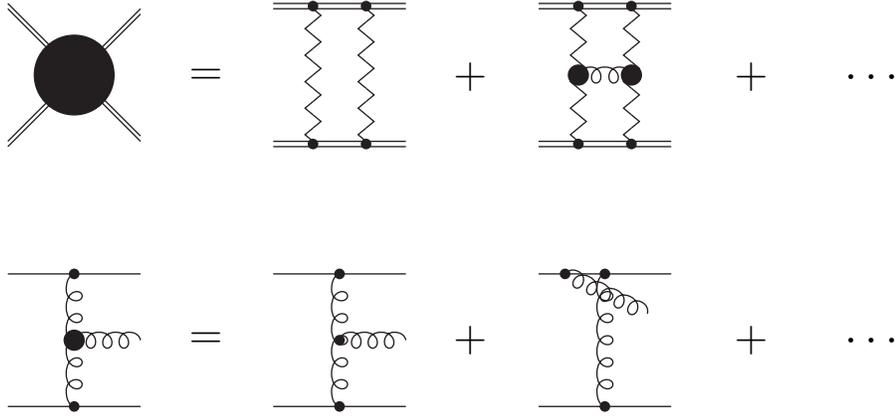
\begin{figure}[htb]

\begin{picture}(400,200)(0,0)

\CCirc(50,150){15}{Black}{Black}
\Line(25,175)(75,125)
\Line(25,177)(75,127)
\Line(25,125)(75,175)
\Line(25,123)(75,173)

\Line(125,175)(175,175)
\Line(125,177)(175,177)
\ZigZag(140,125)(140,175){3}{5}
\ZigZag(160,175)(160,125){3}{5}
\Line(125,125)(175,125)
\Line(125,123)(175,123)
\Vertex(140,124){2}
\Vertex(140,176){2}
\Vertex(160,124){2}
\Vertex(160,176){2}

\Line(225,175)(275,175)
\Line(225,177)(275,177)
\ZigZag(240,125)(240,175){3}{5}
\ZigZag(260,175)(260,125){3}{5}
\Line(225,125)(275,125)
\Line(225,123)(275,123)
\Vertex(240,124){2}
\Vertex(240,176){2}
\Vertex(260,124){2}
\Vertex(260,176){2}
\Vertex(240,150){4}
\Vertex(260,150){4}
\Gluon(240,150)(260,150){3}{2}

\Text(100,150)[c]{{\Large =}}
\Text(200,150)[c]{{\Large +}}
\Text(300,150)[l]{{\Large + $\;\;\;\;\; \cdots$}}

\Line(25,75)(75,75)
\Line(25,25)(75,25)
\Vertex(50,75){2}
\Vertex(50,25){2}
\Gluon(50,25)(50,75){3}{5}
\Vertex(50,50){4}
\Gluon(50,50)(75,50){3}{3}

\Line(125,75)(175,75)
\Line(125,25)(175,25)
\Vertex(150,75){2}
\Vertex(150,25){2}
\Gluon(150,25)(150,75){3}{5}
\Vertex(150,50){2}
\Gluon(150,50)(175,50){3}{3}

\Line(225,75)(275,75)
\Line(225,25)(275,25)
\Vertex(250,75){2}
\Vertex(250,25){2}
\Gluon(250,25)(250,75){3}{5}
\Vertex(235,75){2}
\Gluon(235,75)(266,60){3}{5}

\Text(100,50)[c]{{\Large =}}
\Text(200,50)[c]{{\Large +}}
\Text(300,50)[l]{{\Large + $\;\;\;\;\; \cdots$}}

\end{picture}
\caption{Schematic representation of the total cross section in the 
leading $\ln(s)$ approximation (first row). Double lines represent 
protons or anti-protons, vertical zig-zag lines are Reggeized
gluons, horizontal wavy lines are gluons. The effective vertex for
two Reggeized gluons and one gluon is defined in the second row. Here external
lines can represent quarks or gluons.}
\end{figure}

\begin{figure}[htb]
\begin{center}
{\parbox[t]{5cm}{\epsfysize 12cm \epsffile{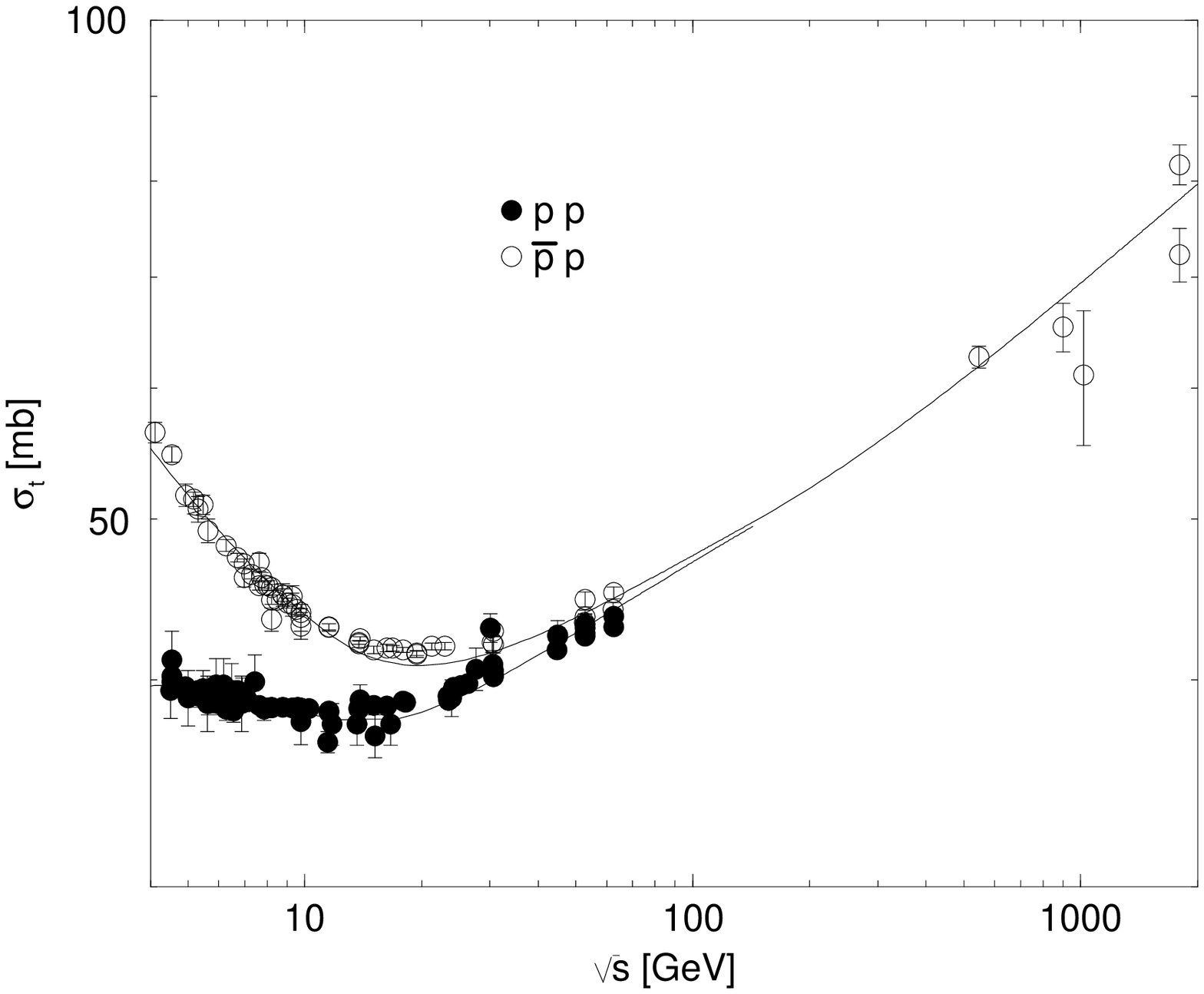}}}
\end{center}
\caption[]{Total cross section calculated up to two gluon rungs and fitted to
the $p \bar p$ and $p p$ data.}
\end{figure}

\begin{figure}[htb]
\begin{center}
{\parbox[t]{5cm}{\epsfysize 12cm \epsffile{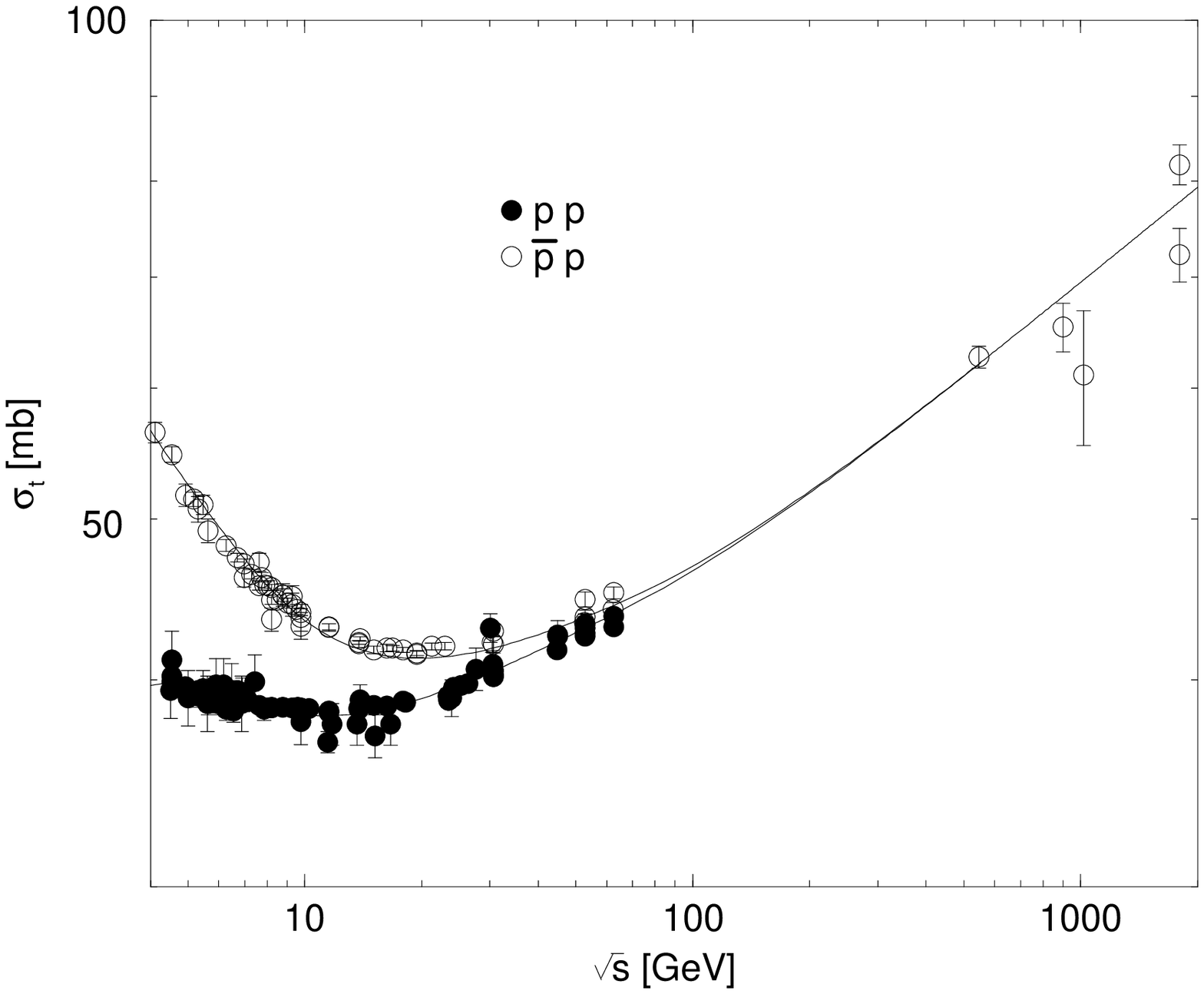} }}
\end{center}
\caption[]{Total cross section calculated up to three gluon rungs and fitted to
the $p \bar p$ and $p p$ data.}
\end{figure}

\newpage

\end{document}